\documentstyle[11pt,aaspp4]{article} 
\begin{document}
\title{Heat conduction in X-ray clusters: Spitzer over 3} 
\author{Andrei Gruzinov}
\affil{Physics Department, New York University, 4 Washington Place, New York, NY 10003}

\begin{abstract}

Effective heat conduction in a random variable magnetic field should be equal to one third of the Spitzer's value. Recent observations indicate that this heat conduction is sufficient to account for the bremsstrahlung in cooling X-ray clusters.

\end{abstract}

\section{Cooling X-ray clusters}

According to observations, some X-ray clusters have bremsstrahlung cooling time much shorter than the Hubble time. These are the so called "cooling flow" clusters (all references). If these clusters live for about a Hubble time in a state similar to their observed state, the radiative energy loss should be compensated for by some energy supply. We will show that the energy might be supplied by heat conduction from large radii. At large radii, the cooling time is longer than the Hubble time and there is enough energy to power the central region for a Hubble time. 

We first show that effective heat conduction in X-ray clusters is equal to one third of the Spitzer's value (\S2). Then we show that Spitzer over 3 possibly does account for all the bremsstrahlung cooling in X-ray clusters listed in our references (\S3), which include all the cooling clusters that we have found in the literature. 

\section{Effective heat conduction in a random time-varying magnetic field}

In an X-ray cluster  plasma, heat conduction along magnetic field is equal to $\kappa \nabla _\parallel T~{\rm erg/cm^2/s}$, were $\nabla _\parallel T$ is the temperature gradient along the field in units of keV/kpc, $\kappa _S=9\times 10^{-4} T^{5/2}$, and $T$ is in keV. In cooling X-ray clusters (all references), in the central 100 kpc region, this heat conduction would cause a noticeable change in a time of order few Gyr. But the cluster plasma is being crisscrossed by about a hundred galaxies on a time scale of order 100kpc/1000km/s=0.1Gyr. On a similar time scale, the magnetic field of the cluster will have to undergo total reconnection, because the galaxies would entrain the field. Thus we have to assume that the magnetic field changes its direction randomly, and  on a short, as compared to the heat conduction, time scale. 

The effective heat conduction in a randomly varying magnetic field is equal to 1/3 of the Spitzer value given above, $\kappa =3\times 10^{-4} T^{5/2}$. This follows, for example, from a formal averaging of the divergence of the heat flux over magnetic field directions. If ${\bf b}$ is the unit vector along the magnetic field, then  
\begin{equation}
<~\nabla \cdot ({\bf b}\kappa _S {\bf b}\cdot \nabla T)~>= (1/3)\nabla \cdot (\kappa _S \nabla T).
\end{equation}

\section{Heat conduction in cooling X-ray clusters}

This author understands neither the origin nor the meaning of the temperature error bars given in the references. The errors were simply ignored in our analysis, and therefore {\it the following procedure does not constitute a real proof} that heat conduction balances the bremsstrahlung in cooling X-ray clusters. But we think that we do have an important clue here.

We will compare bremsstrahlung cooling within the central 100 kpc of all the clusters in our list with heat flowing into the central 100 kps by heat conduction. The bremsstrahlung losses within 100 kpc are calculated as 
\begin{equation}
B=4\pi r^3\epsilon,
\end{equation}
where $\epsilon =6.5\times 10^{-24}n^2T^{1/2} {\rm erg/cm^3/s}$ is the cooling rate at 100kpc, with $n$ in ${\rm cm^{-3}}$ and $T$ in keV measured at 100 kpc.  This formula would be exact for $n \propto r^{-1}$ and $T=const$; both assumptions are approximately correct in the vicinity of $r=100$kpc. 

To calculate the heat inflow due to heat conduction we need temperature gradients, and measuring those using the noisy graphs found in the references is not the right procedure. Instead we have selected two best clusters (or maybe the best deprojection techniques ?) that give the smallest error bars near 100kps -- Abell 1795 and Abell 2199. Ettori et al (2001) give the temperature fitting formula $T\propto r^{0.27}$, thus the temperature gradient is  simply $dT/dr=0.27T/r$. On the other hand, Fig. 6 of Johnstone et al (2002) can be fitted by $T=1+1.5\log r$, with $T$ in keV and $r$ in kpc. At 100 kpc this gives $dT/dr=0.38T/r$. We will assume that at 100 kpc, $dT/dr=(1/3)T/r$ for all clusters. Then the heat flux into the central 100 kpc is
\begin{equation}
H=(4/3)\pi r\kappa T.
\end{equation}

The ratio of heat influx and bremsstrahlung cooling at 100 kpc is then 
\begin{equation}
{H \over B}={0.005T^3\over r^2 n^2},
\end{equation}
where $T$ is in keV, $r$ in kpc, $n$ in ${\rm cm^{-3}}$. The table lists the density and temperature at 100 kpc in cooling X-ray clusters and the calculated ratio $H/B$. We have ``measured'' $n$ at 100 kpc and $T$ at 100 kpc from the graphs in all the references. Hydra A comes two times because the corresponding graph gives two temperatures. In all these clusters, at about 100 kpc, where the cooling time becomes shorter than the Hubble time, the heat influx from the outside approximately balances the bremsstrahlung cooling inside.

\begin{tabular}{lccr}

Cluster & $n$ at 100 kpc, $0.01{\rm cm^{-3}}$  & $T$ at 100 kpc, keV & ~~~~~~~~~~~~$H/B$ \\
& & & \\
3C295 & 0.9 & 5.0 & 0.8 \\
RXJ1347.5-1145 & 3 & 11 & 0.7 \\ 
Abell 2052 & 0.4 & 3.5 & 1.3 \\ 
Hydra A & 0.5 & 3.65 & 1.0 \\ 
Hydra A & 0.5 & 4.1 & 1.4 \\ 
Abell 1795 & 1.1 & 5.1 & 0.5 \\ 
Abell 2199 & 0.6 & 4.0 & 0.9 \\ 
Abell 2597 & 0.6 & 3.3 & 0.5 \\ 
Abell 1835 & 1.8 & 7.7 & 0.7 \\ 

\end{tabular}

\acknowledgements

I thank John Bahcall, Patrick Diamond, and Ramesh Narayan for useful discussions.

\end{document}